\documentclass{aastex}          
\usepackage{spr-astr-addons}    
\usepackage{natbib}
\usepackage{url}
\usepackage{amsmath}
\begin{document}
\title{Ghost Collapse : exploring feasibility of spurious Spherical Collapses}
\shorttitle{Spurious cases of Spherical Collapse}

\author{Aditya Vidhate\altaffilmark{1}} 
\and 
\author{Rahul Nigam\altaffilmark{1}}

\altaffiltext{1}{BITS Pilani Hyderabad Campus}

\begin{abstract}
We explore the real solutions to the Spherical Collapse Model in a non-flat Universe with a Cosmological Constant, and observe a possible situation for a fake or Ghost Collapse, in which an expanding overdense spherical region, turns around and begins to collapse, turns around again after a finite time and starts expanding. To make such a situation of spurious collapse feasible, we make a linear redshift dependent correction to the standard Dark Energy density term which is originally in the form of a cosmological constant. There is good reason to believe in such a correction based on recent research which hints that Dark Energy desnity evolves with the redshift (even becomes negative) when fit to observational data.

\end{abstract}


\section{Introduction}
The Universe is fairly homogenous and isotropic on large scales \cite{AD}, however, the existence of vast variety of structures on smaller scales gives a true glimpse of the inhomogeinity of the Universe. All the structure in Universe is made up of bound mass, an example of Gravity winning over the expansion of our cosmos. The easiest and most basic approach to understand this structure formation at first approximation is given by the Spherical Collapse Model \cite{GJ}. The model considers an expanding overdense spherical region, and finds the conditions for it to turnaround and start collapsing to form a bound massive object. A single turnaround point to the Spherical Collapse model, as given by the equation describing the evolution of the gravitating spherical region  has been studied in great detail. In this research we focus on all the real roots of the aforementiond equation \ref{the cubic} to explore cases of multiple chains of turn arounds, creating situation wherein a collapse doesn't guarantee consistency till a bound object is formed, but is rather prone to a second turnaround, leading to an expansion.

For such a Ghost Collapse situation to occur, the Universe needs to evolve in a peculiar way, at least in the local environment, especially requiring an evolving Dark Energy density. The idea for the solution to this comes from recent studies which hint at an evolving Dark Energy density as studied by Wang et.al. \cite{WY} who reconstruct evolving Dark Energy using observations and Dutta et.al. \cite{DK}, who perform a model independent analysis of the Dark Energy density hinting at its evolving nature and also the confident possibility of the Cosmological Constant being negative. These papers break the long known constant nature of the cosmological constant, and put a rather interesting evolving dark energy density playing on top of the cosmological constant (at the minimum). We use this inspiration to devise a certain correction to the Cosmological Constant, that makes it evolve in time, to explore situations in which the Ghost Collapse becomes more feasible.

The following sections will build onto these two basic ideas - the cases of 3 real roots to Spherical Collapse model and the correction in Dark Energy density evolution. Section \ref{sphcol} contains all relevant equations used to reach the turnaround equation and the necessary computations done. Section \ref{decorrection} contains details of the Dark Energy corrections. Section \ref{method} will deal with the technical details of how the equations were solved numerically. Section \ref{result} will showcase the data on the basis of which we built the model. Possible future work in this direction is outlined in Section \ref{future}.

\section{Spherical Collapse}\label{sphcol}
The Spherical Collapse model is a very well known and vastly studied model describing the formation of bound objects by evolution of uniformly overdense spherical regions in a smooth background matter density. Owing to the nobel prize winning observations of S. Perlmutter, A. Riess and B. Schmidt discovering an acceleratingly expanding universe, implying a pressing need for Dark Energy, possibly manifesting in the form of the famous Cosmological Constant, the Spherical Collapse model required modifications to incorporate these effects into it. A concise set of equations for the Spherical Collapse Model in presence of a cosmological constant in the FLRW metric is presented by Lokas et. al.\cite{LE}, the evolution happening in a non-flat Universe. The set of equations, including all the relevant ones required to understand them, have been listed below.

\subsection{Relevant Equations}
The Friedmann equations for a non-flat Universe having an FLRW Metric with a cosmological constant is given as follows
\begin{equation}
\left( \frac{da}{dt}\right)= H_0\left[ 1+\Omega_0\left(\frac{1}{a}-1\right) + \lambda_0\left( a^2-1\right)  \right] ^{1/2}
\end{equation}

Where $H$ is the Hubble parameter, $\Omega$ is the Matter density parameter, $\lambda$ is the Dark Energy desnity parameter, $a$ is the scale factor of the universe normalised to 1 at present time and all the variables with subscript $0$ are the present values of the parameter.

The evolution of Dark Energy and Matter is respectively governed as follows (where the cosmological constant and Matter are both expressed as density parameters i.e. normalised with critical density)

\begin{equation} \label{L evltn}
\lambda(z)=\lambda_0\left[\frac{H_0}{H(z)}\right]^2
\end{equation}

\begin{equation} \label{O evltn}
\Omega(z)=\Omega_0(1+z)^3\left[\frac{H_0}{H(z)}\right]^2
\end{equation}

Both the above are governed by the cosmology that describes Hubble parameter evolution as follows
\begin{equation} \label{Hubble evltn}
\left[\frac{H(z)}{H_0}\right]^2 = \Omega_0(1+z)^3 - (\Omega_0 + \lambda_0 -1)(1+z)^2 + \lambda_0
\end{equation}

The evolution of a Spherically Overdense region can be found at any time $t_i$, having an overdensity $\Delta_i$ over the background density and of radius $r_i$. Using the variable $s=r/r_i$, it is given as follows
\begin{equation}\label{derv}
\frac{ds}{dt} = \frac{H_{i}}{g(a,s)}
\end{equation}

Where, the subscript $i$ denotes the values of the parameters at a redshift $z_i$, and $g(a,s)$ is as follows
\begin{equation}
g(s)=\left[ 1+\Omega_{i}(1+\Delta_{i})\left(\frac{1}{s} -1 \right) + \lambda_{i}(s^{2}-1) \right]^{-1/2}
\end{equation}

Using the conservation of energy inside the Spherical Region, the turnaround radius $r_{ta}$ ($s_{ta}=r_{ta}/r_i$) is found by taking the derivative (eq:\ref{derv}) to zero. The solution is given in the form of the cubic equation as follows
\begin{equation} \label{the cubic}
b_{1}s_{ta}^{3} + b_{2}s_{ta} + b_{3} = 0
\end{equation}

Where, \\
$ b_{1} = \lambda_{i}$\\
$ b_{2} = 1-\Omega_{i}(1+\Delta_{i}) - \lambda_{i}$\\
$ b_{3} = \Omega_{i}(1+\Delta_{i})$\\

All the parameters of the cubic equation are made up of 3 adjustable terms that depend on the Cosmology being considered; $\lambda$ (density parameter corresponding to Dark Energy), $\Omega$ (density parameter corresponding to the matter contained in the Universe) and $\Delta$ (the matter overdensity in a spherical region, over the background density).

\subsection{Roots to the turnaround equation}
We clearly can have a maximum of 3 real roots the turnaround equation described by equation \ref{the cubic}, owing to its cubic order. Specifically focusing on cases with 3 real roots, we can directly conclude that at least one and at most two roots will be negative as the sqaure ordered term in equation \ref{the cubic} is missing, implying the sum of roots to be exactly zero.\\

We numerically iterate various values of the set $\{\lambda,\Omega,\Delta\}$, replace them into the turnaround equation and find roots for each instance. The details of the method are given in the methodology section. From all the roots that we find for each set, we segregate the cases of 3 real roots. From all the 3 real root cases, we further segregate cases with 2 positive real roots.\\

Finally, we have a dataset of many cases, each with a unique value for the set $\{\lambda,\Omega,\Delta\}$, such that the turnaround roots have 2 positive real values, out of the 3 real values. The 2 roots correspond to a collapse root - where the spherical region expands, turns around and collapses; and the expansion root - where the spherical region while collapsing, turns around, and expands. Such cases are interesting to probe into, to figure out if a situation for a fake Collapse, or in our terminology, Ghost Collapse, can happen.

\section{Dark Energy Corrections}\label{decorrection}
\subsection{Ghost Collapse}

As setup in the previous section, the solutions to spherical collapse hint at a possible situation occurring wherein a turnaround happens, and a spherical overdensity begins to collapse, leading on to face another turnaround, after which the overdensity keeps expanding. The cartoon sketch below illustrates the situation in which we are interested and to study such phenomenon, which we refer to as Ghost Collapse, we analyse the second derivatives of the energy conservation equation (which is the first derivative of the turnaround equation \ref{the cubic}).\\

\begin{figure}[t!]
\includegraphics[width=0.5\textwidth]{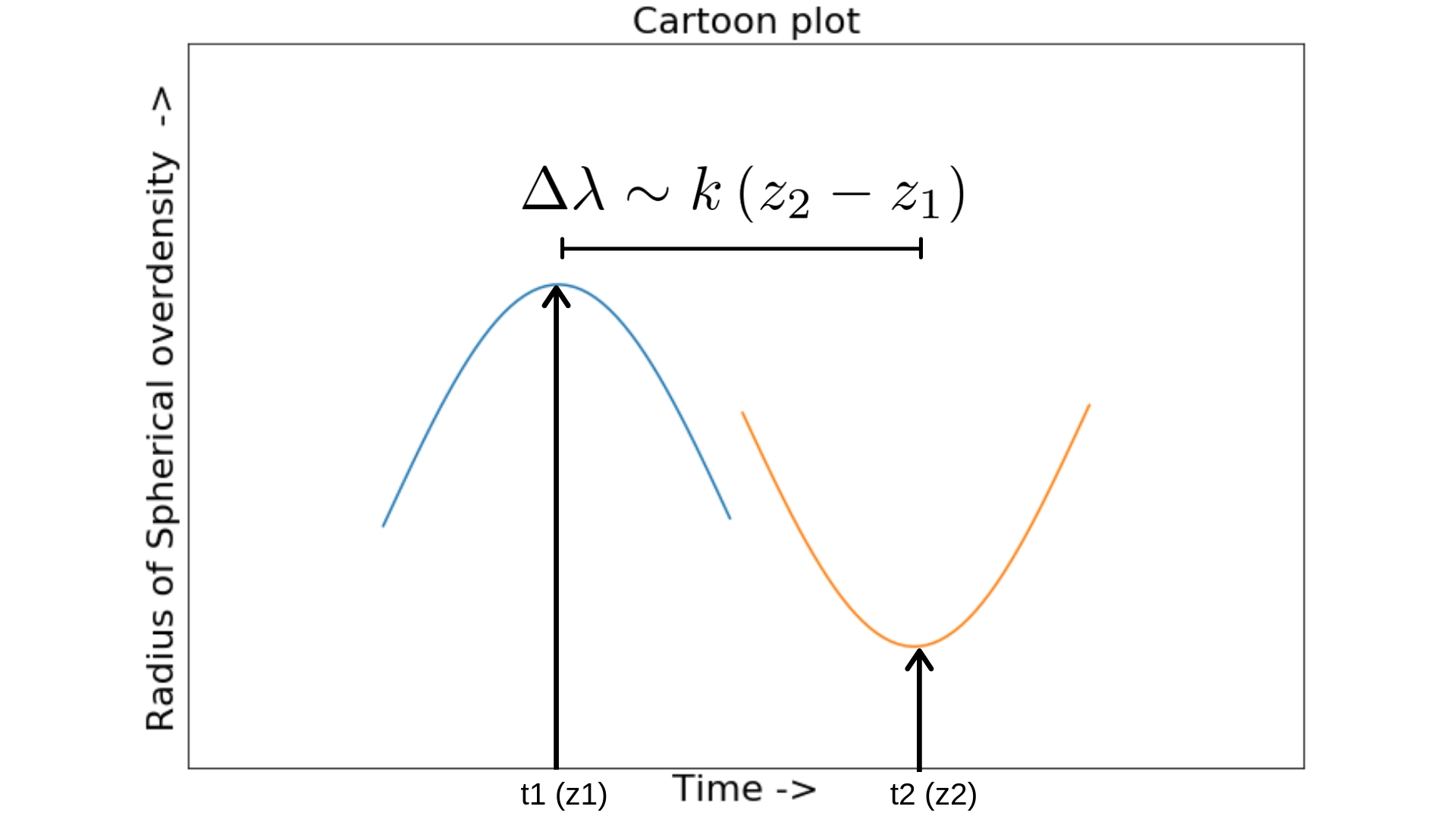}
\caption[The radii at which two turn arounds happen]{The structure initially follows the blue line for a given $\lambda$ and turns around at a maximum to collapse. At this point $\lambda$ changes to a different value for which structure now follows maroon line and turns around again but now to expand forever.}
\label{gcollapse}
\end{figure}

This analysis leads to datasets of the type given below. A common theme is observed with the 2 positive roots from cases as compiled in the previous section (the negative one being non-physical) - the smaller root for turnaround leads to a collapse (negative 2nd derivative) and the bigger root for turnaround leads to an expansion (positive 2nd derivative).\\

For a Ghost Collapse to happen, the smaller root (collapse) should be encountered after the bigger root (expansion). More explicitly, if the smaller root - $s_{coll}$ (corresponsing to a collapse), of equation \ref{the cubic} with the characterising parameters being $\{\lambda_{coll},\Omega_{coll},\Delta_{coll}\}$, is bigger than the bigger root - $s_{exp}$ (corresponsing to an expansion), of equation \ref{the cubic} with different characterising parameters $\{\lambda_{exp},\Omega_{exp},\Delta_{exp}\}$, then a situation of Ghost Collapse becomes feasible as it matches the description given above.
From hereon, we develop a Toy Model for the Universe in which such situations are feasible and we go about this with one key change.

\subsection{Modifications to Dark Energy}
The paper by Dutta et.al.\cite{DK} and Wang et.al. \cite{WY} hint that according to observations the Dark Enrgy density is evolving. This opens up a very wide perspective on how the Dark Energy can behave, and shifts the perspective from a cosmological \textit{constant}. We use this recent development to construct a scenario where the Dark Energy behaves as a small linear redshift dependent correction to the baseline cosmological constant, as described in the formula below,
\begin{equation} \label{z correction}
\lambda=\frac{\lambda_0}{H^{2}}(1+K(z_2-z_1))
\end{equation}\\

Where, $K$ is a small constant found from simulated data, $z_1$ and $z_2$ are the redshifts at which the first and second turnarounds respectively happen.

This particular form for the correction is chosen for its simplicity, without any specific mathematical implications. The goal to find a better functional dependence on $\delta z=z_2-z_1$ will form the basis of future work.

We assume that the FLRW metric for the Universe is still consistent as the correction to the cosmological term is small. This also implies that the correction term in Dark Energy doesn't change the Energy Conservation equation (as it is derived from the Friedmann Equations only).
\\
The matter density evolves in the same way as it does in an FLRW Universe.
\begin{equation} \label{omega evltn}
\Omega \sim (1+z)^3
\end{equation}

Because the background density evolves as \ref{omega evltn} and assuming that the overdensity in any confined region in space stays constant in proper space, it evolves as
\begin{equation}
\Delta \sim \frac{1}{(1+z)^3}
\end{equation}

This also implies that the ratio of the background density to that of overdensity always remains constant
\begin{equation} \label{consistency}
\frac{\Delta}{\Omega} \sim Constant
\end{equation}

Using the consistency equation \ref{consistency}, we put a constraint on the feasible solutions from our dataset and filter the passable data points out.
\\
By using equation \ref{z correction} and equation \ref{O evltn} we get the correction term for Dark Energy in terms of the the redshifts $z_1$ and $z_2$, where $z_1$ corresponds to redshift at which the collapse begins and $z_2$ refers to the redshift at which the second turnaround begins.
\\
For various values of $z_1$ and $z_2$, we get different values for the correction term, as described elaborately in the Methodology section. We need the correction term to be very small to not affect the FLRW metric and the energy conservation equations and we need the difference between redshifts to be feasible. Using all such constraints we narrow down on the datasets as presented below.

\begin{figure*}[t!]
\centering
\includegraphics[width=0.8\linewidth]{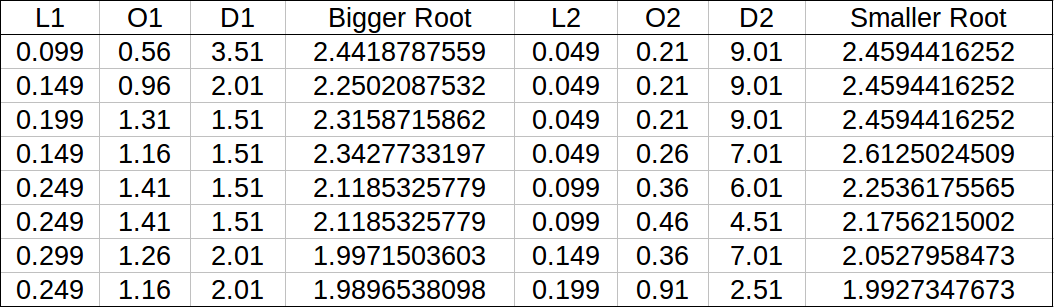}
\caption[The radii at which two collapses happen]{Observe that the smaller radius is larger than the larger radius. The smaller root is the one at which the real collapse happend. The larger root is the one at which the collapse turnsaround and begins to expand again.}
\label{ghosts}
\end{figure*}

\section{Methodology}\label{method}
\subsection{Numerical roots to turnaround equation}
The parameters of the turnaround equation \ref{the cubic} is constructed by numerically varying the values of $\lambda$,$\Omega$ and $\Delta$ with steps as described below
\begin{itemize}
\item $\lambda$ was varied with steps 0.05, starting from value -0.5 to value 1.0.

\item $\Omega$ was varied with steps 0.05, starting from value 0.0 to value 1.5.

\item $\Delta$ was varied with steps 0.5, starting from value -1.0 to value 10.0.
\end{itemize}

\subsection{Calculations to Corrections of Dark Energy}
The correction to the Dark Energy is in the form of an added term that depends on the difference in redshifts of 2 turnarounds multiplied by a constant (K), as can be concluded from the equation for evolution of $\lambda$ as given in equation \ref{L evltn}.
\begin{equation}
\lambda(z_2)=\lambda(z_1)\left[\frac{H(z_1)}{H(z_2)}\right]^2 \left(1+K(z_2-z_1) \right) 
\end{equation}
In the limit K tends to 0 for a small correction, the equation returns to the original form.\\
Rearranging the equation as follows,
\begin{equation} \label{crctn 1}
\frac{\lambda(z_2)}{\lambda(z_1)}=\left[\frac{H(z_1)}{H(z_2)}\right]^2 \left(1+K(z_2-z_1) \right) 
\end{equation}

From the equation for evolution of $\Omega$ as given in equation \ref{O evltn}, we have
\begin{equation} \label{crctn 2}
\frac{\Omega(z_2)}{\Omega(z_1)}=\frac{(1+z_2)^3}{(1+z_1)^3}\left[\frac{H(z_1)}{H(z_2)}\right]^2
\end{equation}

Dividing equation \ref{crctn 1} with \ref{crctn 2} we get
\begin{equation}
\frac{\lambda(z_2)}{\lambda(z_1)}\frac{\Omega(z_1)}{\Omega(z_2)}=\frac{(1+z_1)^3}{(1+z_2)^3}\left(1+K(z_2-z_1) \right) 
\end{equation}

Rearranging the above equation have only $K$ on Right Hand Side,
\begin{equation}
\left[\frac{\lambda(z_2)}{\lambda(z_1)}\frac{\Omega(z_1)}{\Omega(z_2)}\frac{(1+z_2)^3}{(1+z_1)^3}-1\right] \frac{1}{(z_2-z_1)}=K
\end{equation}

We now do a brute force iteration over many values of $z_1$ and $z_2$, substituting them in the equation for each pair $\{z_1,z_2\}$ and getting a value of $K$. From all such values of $K$s we choose the ones that are small and fit our physical description feasibly.

\section{Results}\label{result}
Figure 1 shows identified cases of Ghost Collapses from the analysis described in sections above. Each line presents a pair of turnarounds in different environments having $\lambda$, $\Omega$ and $\Delta$ as $\{L1,O1,D1\}$ and $\{L2,O2,D2\}$. The values of the root under the title ``\textit{Smaller Root}'' corresponds to a turn aroound that leads to a real collapse. The values under the title ``\textit{Bigger Root}'' corresponds to a turn around that leads to expansion of the spherical region. As can be clearly observed, the values under \textit{Smaller Root} are larger than the values under \textit{Bigger Root} implying that the expanding turn around can happen after the collapsing turn around has happened.

The larger the redshift $z$, the more in the past an event is. The results from the numerical analysis for the redshift dependences of the Ghost Collapse turn arounds are shown in Figure 2, wherein we get many ordered pairs of redshifts for each turn around pair in Figure 1, the smaller redshift corresponding to the collapsing turn around and the larger redshift corresponding to the expanding turn around as it happens \textit{after} the initial collapase. These redshift dependences give us a very important estimate on the value of $K$ - the correction term to Dark Energy density as described in eq:\ref{z correction}.

\section{Summary}\label{summary}
\begin{itemize}
\item The cubic equation \ref{the cubic} for non-flat universe with a cosmological constant, describing collapse of a spherically overdense region is used to find all roots which correspond to turnaround radius of the regions using brute force numerical substitution of $\Lambda$, $\Omega$ and $\Delta$. All cases with 3 real roots are found.

\item From these we filter all cases with 2 positive real roots (1 root has to be negative to keep the squared term 0 of the cubic). We observe that the real collapse happens at the smaller root (smaller radius) (negative second derivative) and the blow up happens at the larger root (larger radius).

\item We filter cases in which the bigger root ($r_{exp}$) is smaller than the smaller root ($r_{col}$). Such cases imply that the spherical region first undergoes a collapse and then at a smaller radius turns around again to blow up. We call this the Ghost Collapse situation.

\item We try to make such an event consistent by adding a correction term to the dark energy which allows such a consecutive turnaround to happen. This puts a condition on the consistency of overdensity and the matter density. Using this constraint further filtering is done on the roots.

\item From thereon we finally try to conclude the appropriate correction term which can make such a situation feasible.
\end{itemize}

\section{Conclusions}\label{conclu}
The possibility of a Ghost Collapse as explored in the paper, is undoubtedly exciting and has the potential to impact structure formation as we know it. Most of the previous work in studying Spherical Collapse in a presence of Dark Energy has been constrained by taking Dark Energy as a Cosmologocal Constant and observing the changes in the critical overdensity for collapse. With our study, we have consistently aimed at pushing beyond the previous perspectives.\\

Although the scenarios discussed in the paper are mere possibilities, it drives forward the much more important idea of an evolving Dark Energy density in our Universe, which if established, will give birth to a vast variety of many exotic phenomena such as Ghost Collapses.\\

\section{Future Work}\label{future}
The possibilities discussed in this paper gives birth to many opportunities for research and exploration, some of which we briefly discuss in this section.\\

One of the most exciting projects would be to modify Cosmological N-Body simulation codes to imitate a local evolving Dark Energy density, and to observe if Ghost Collapses can actually happen. Such experiments can give much more insight into the evolution of such exotic phenomenon, and can be especially used to quantify the processes in non-linear regimes. Such studies would also help in quantifying the probability with which Ghost Collapses can occur in the Universe.\\

Another important project would be check if a better correction to the Dark Energy Density be found, which is more effective and feasible in our observable cosmos.\\

\acknowledgments
We would like to thank BITS Pilani Hyderabad Campus to provide all the infrastructure necessary to carry out this work.

\end{document}